\documentclass[a4paper,USenglish]{lipics-v2021}
\usepackage{amsmath}
\usepackage{multirow}
\usepackage{multicol}
\usepackage{subcaption}
\usepackage{amsthm}
\usepackage{relsize}
\newcommand{\ie}{\emph{i.e.} }
\newcommand{\eg}{\emph{e.g.} }

\usepackage{xspace}

\newcommand{\macro}[2]{ \providecommand{#1}{{\ensuremath{#2}}\xspace}}

\macro{\N}{\mathbb N}
\macro{\R}{\mathbb{R}}
\macro{\dG}{\bf G}
\macro{\dH}{\bf H}
\macro{\dK}{\bf K}
\macro{\dfam}{\mathcal F}
\macro{\algo}{\mathcal A}
\macro{\proc}{\odot}

\newcommand{\macromath}[2]{\providecommand{#1}{{{\ensuremath{#2}}}\xspace}}
\newcommand{\rmacromath}[2]{\renewcommand{#1}{{{\ensuremath{#2}}}\xspace}}

\rmacromath{\b}{\circ}
\macromath{\n}{\bullet}
\macromath{\g}{\textcolor{gray}{\bullet}}
\macromath{\bn}{\b\!\!-\!\!\!\!\!\!-\!\!\n}
\macromath{\bg}{\b\!\!-\!\!\!\!\!\!-\!\!\g}
\macromath{\gn}{\g\!\!-\!\!\!\!\!\!-\!\!\n}
\macromath{\lnoir}{\b\!\!\rightarrow\!\!\n}
\macromath{\lblanc}{\b\!\!\leftarrow\!\!\n}
\macromath{\lall}{\b~~ \n}
\macromath{\lnall}{\b\phantom{\!\!-\!\!\!\!-\!\!}\n}
\macromath{\lok}{\b\!\!\leftrightarrow\!\!\n}
\macromath{\eps}{\varepsilon}

\macromath{\I}{\mathcal I}
\macromath{\Out}{\mathcal O}
\macromath{\ma}{\mathcal M}

\usepackage{tikz}
\usepgfmodule{shapes}
\usetikzlibrary{calc,
                positioning,
                patterns,
                topaths,
                arrows.meta,
                decorations.pathreplacing}

\tikzset{p/.style={circle, draw, fill=black!50,
                        inner sep=0pt, minimum width=6pt}}

\tikzstyle{proc}=[circle, draw, inner sep=0pt,
                  minimum width=8pt, text width=4pt, line width=0.1pt]
\tikzstyle{proc1}=[circle, draw, inner sep=0pt,
                  minimum width=5pt, line width=0.1pt]
\tikzstyle{proc2}=[circle, draw, inner sep=0pt,
                  minimum width=3pt, line width=0.1pt]
\tikzstyle{proc3}=[circle, draw, inner sep=0pt,
                  minimum width=1pt, line width=0.1pt]
\tikzstyle{rel}=[>=stealth]  
\tikzstyle{b}=[fill=white]
\tikzstyle{g}=[fill=gray]
\tikzstyle{n}=[fill=black]

\macro{\proc}{p}
\macro{\chr}{\mathrm Chr}
\newcommand{\bigwr}{{\mathlarger{\mathlarger{\wr}}}}
\DeclareMathOperator*{\causal}{\rightsquigarrow}

\usepackage{enumerate}
\usepackage[vlined,linesnumbered,ruled]{algorithm2e}
\usepackage{tikz} %

\usepackage{fancyhdr}
\fancypagestyle{plain}{%
  \fancyhf{} %
  \fancyfoot[R]{{{\thepage}}} %
}
\fancypagestyle{empty}{%
   \fancyhf{} %

 }

\title{A Topology by Geometrization for Sub-Iterated Immediate Snapshot Message Adversaries and Applications to Set-Agreement}
\author{Yannis Coutouly}{Laboratoire d'Informatique et Systèmes - Université Aix-Marseille \and CNRS}{}{}{}
\author{Emmanuel Godard}{Laboratoire d'Informatique et Systèmes - Université Aix-Marseille \and CNRS}{}{}{}

\nolinenumbers
\begin{document}

\maketitle

\begin{abstract}
  The Iterated Immediate Snapshot model (IIS) is a central model in
  the message adversary setting.  We consider general message
  adversaries whose executions are arbitrary subsets of the executions
  of the IIS message adversary.  We present a
  new topological approach for such general adversaries, based upon
  geometric simplicial complexes.
  We are able to define a topology directly on the considered sets
  of executions, which gives both simpler and more powerful ways of
  using topology for distributed computability.

  As application of this new framework, we present a complete characterization and
  lower bounds for solving set-agreement for general sub-IIS
  message adversaries. 
\end{abstract}

\section{Introduction}

Since the seminal works of Herlihy-Shavit, Borowsky-Gafni and Saks
Zaharoglou \cite{HS99,SZ,BoGa93}, using topological methods has proved very fruitful for
distributed computing and for distributed computability in
particular.
Since those first results, the topological framework has been refined to be
presented in a more effective way. In particular, the Iterated Immediate Snapshot
model ($IIS_n$)  is a special message adversary that has been proposed
as a central model to investigate distributed computability.
In this note, we propose an enhanced presentation of the
topological approach in order to investigate distributed computability and complexity for message
adversaries more general  than $IIS_n$.

The Iterated Write Snapshot is a shared memory model. Single-writers/multi-readers registers are
accessible by processes. There is usually as many registers as
processes and the registers are arranged in a one-shot array.
It can be assumed, as in \cite{BGatomicsnapshot}, that there is a
\texttt{writeSnapshot()} primitive that enables processes to atomically write
values to their register and read the
content of the full array. Each concurrent access can read the values
corresponding to the calling process and also the values previously written by
other processes in the same round.
In a given round, all possible interleaving of calls to
\texttt{writeSnapshot()} are allowed.

The main interest of this model is that it has a simple synchronous
and regular structure and that it was proved in \cite{BGequivIIS} that
a bounded colourless task can be wait-free solved in the classical read-write
model if and only if it is solvable in the Iterated Write Snapshot model.
So this model has the same computing power as shared memory, but using
topological tools is
simpler in this model (see the tutorial in \cite{DCcolumn} and the
thorough coverage of \cite{HKRbook}).

\medskip

Later, the Iterated Immediate Snapshot model has been introduced which
is the counterpart of Iterated Write Snapshot model to message
passing.
Process may never fail, however, it is possible that a ``fast'' correct process
never sees another correct process.
The communication structure that one gets with a standard shared memory model
is usually edge-transitive : in a round, if a process $p$ ``sees`` a
process $p'$ that also ``sees`` a process $q$, then $p$ also ``sees``
$q$.
Considering message passing systems, 
this condition is actually not necessary, and in
\cite{messadv}, where the terminology of ``message adversaries'' is
introduced, this condition is dropped by considering various families
of directed graphs where the instant  graphs are not transitive.
In \cite{messadv}, Afek and Gafni show that 
the same tasks can be solved as in standard asynchronous shared memory
if  the instant directed graphs are tournaments.
So some message adversaries with specific non-complete underlying graph of communication
can have the same computing power as classical shared memory.

Subsequently, Raynal and Stainer have shown in \cite{RSequiv-conf}
that it is possible to consider various message adversaries (they are
not all iterated) where further restrictions on the set of possible
scenarios, \ie weaker adversaries, correspond to well known
asynchronous shared-memory models enriched with failures detectors.
It appears that the general message adversary model is a very rich,
and at some time very convenient (some message adversaries have very simple protocol
complex) model to describe distributed systems from the point of view
of computability. It is therefore of great interest to investigate as
general as possible message adversaries, since these can \emph{model} numerous type of
failures while being actually fault-free by themselves.

\medskip
We consider the setting of Iterated Immediate Snapshot message
adversary ($IIS_n$), a system of $n+1$ processes whose messages 
exchanges satisfy the Immediacy and Containment properties (see
Section~\ref{sect:iis}).
In this paper, we investigate arbitrary sub-IIS message adversaries,
that is message adversaries whose set of executions are arbitrary
subsets of the executions of the $IIS_n$ message adversary.
We present a new topological method that enables
to define a topology directly on the set of executions of $IIS_n$, for
any $n$. We use the precise description we give of this new topology
to give a full characterization of sub-IIS message adversaries solving
the classical set-agreement problem.

\subsection{Motivation and Contributions}

In order to correctly handle general message adversaries,  
contrary to the usual focus in topological methods for
distributed computing, we consider simplicial complexes primarily as \emph{geometric}
simplicial complexes.
We show then that it is possible to associate, via a natural $geo$ mapping,
any infinite execution of $IIS_n$ to a point of the standard euclidean
space $\R^N$.
The topology on the set of executions is then the topology induced
from the standard topology by the mapping $geo$~: the open sets are
pre-images $geo^{-1}(\Omega)$ of the open sets $\Omega$ of $R^N$.
The standard euclidean topology is simple and well understood, however, since $geo$
is not injective, it is necessary to describe so-called
``non-separable sets'' in order to fully understand the new topology.
In topology, two distinct elements
$x,y$ are said to be non-separable if for any two neighbourhoods
$\Omega_x$ of $x$ and $\Omega_y$ of $y$, we have
$\Omega_x\cap\Omega_y\neq\emptyset$. In our setting, two executions
are non-separable when they have the same image via the mapping
$geo$. So in Section~\ref{sec:geoequiv}, we first investigate
$geo$ pre-image sets of given points in $\R^N$.

\medskip
The standard chromatic subdivision is the combinatorial topology
representation of one round of the Immediate Snapshot model. Its simple
and regular structure makes topological reasoning attractive.
In this paper, we introduce a new universal algorithm and show its
simple relationship with the standard chromatic subdivision as exposed
in the geometric simplicial complex setting. This new algorithm
averages with specific weights vectors of $\R^N$ at each node. We
denote this averaging algorithm the Chromatic Average Algorithm. 
Running the Chromatic Average Algorithm in the $IIS_n$ model gives a natural
geometric counterpart in $\R^N$ to any given execution of $IIS_n$. The mapping
associating executions of $IIS_n$ to points in $R^N$ is the geometrization $geo$.

We present in Theorem~\ref{thm:geoEquiv} a complete combinatorial
description of the different kind of pre-image sets $geo^{-1}(x)$ with
$x\in\R^N$,
so it describes all non-separable sets in $IIS_n$.  Interestingly,
we show that they are of only three possible size : 1, 2 and infinite size.

Since we do have non-separable sets in our setting, it shows that the
standard abstract simplicial complexes approach is actually not always
directly usable, since abstract simplicial complexes are known to have
separable topology. It means that, for the first time, we could have
to explicitly use the geometric version of simplicial complexes to
fully investigate general distributed computability, in particular for
non-iterated message adversaries. We call the topology defined here
the \emph{geometrization topology} to emphasize this change of
paradigm.

\medskip
The $k-$set-agreement problem is a standard problem in distributed
computing and it is known to be a good benchmark for topological
approaches. The $k-$set-agreement problem is a distributed task where processes have to
agree on no more than $k$ different initial values.  The set-agreement
problem is the $k-$set agreement problem with $k+1$ processes.
In the shared memory model, the impossibility of wait-free $k-$set
agreement for more than $k+1$ processes is one of the crowning
achievements of topological methods in distributed computing
\cite{HS99,SZ,BoGa93}.

We apply our technique to derive a characterization and lower bounds
for general message adversaries solving set-agreement.
The characterization of Th.~\ref{th:carac} states that set-agreement is solvable for
$\ma\subset IIS_n$ if and only if the geometrization of \ma has an
``hole'', \ie $geo(\ma)$ does not cover the convex hull of $S^n$, the 
simplex of dimension $n$.

This new topological approach enables to efficiently handle non-iterated message
adversaries in a general way for the first time. As is seen
in the application to set-agreement, it is possibly to investigate new
phenomenon. Moreover, in difference to what is usually implied,
it seems that handling general adversaries is not really a question of compacity
but way more a question of separability. Here, the characterization we
give is also appropriate and directly applicable to non-compact adversaries.

\subsection{Related Works}
We have shown in the previous section that the model considered here is very relevant in many ways.  This
model of synchronous communication has actually been introduced
numerous times under different names.  We mention briefly the mobile
omissions model \cite{timeisnotahealer}, the more recent
``Heard-of'' model \cite{CS09}, the iterated snapshot model
\cite{BGatomicsnapshot} and its final evolution as a \textit{message
  adversary} model \cite{messadv}.  Some equivalences have been proved
between these synchronous presentations and asynchronous models in the
case of colourless tasks \cite{BGequivIIS}.  Note also that in the
case of dynamic networks, whenever the communication primitive is a
broadcast (to the current neighbours), this model can also be used.

In \cite{GKM14}, Gafni, Kuznetsov and Manolescu investigate models that are subsets of the
Iterated Immediate Snapshot model. They introduced \textit{ad hoc} ancillary
infinite simplicial complexes, that are called \emph{terminating
  subdivisions}. We believe our tools can provide a simpler, and
less error-prone, framework to investigate distributed computability
of sub-IIS models.

In a series of works, averaging algorithms to solve relaxed versions of the
Consensus problem, including approximate Consensus, have been
investigated. In \cite{approxAverage}, Charron-Bost, Függer, and  Nowak have
used matrix oriented approaches to show the convergence of different averaging algorithms.
We use a similar stochastic matrix technique here to prove the convergence of the
Chromatic Average Algorithm. In \cite{FNS21}, Függer, Nowak and Schwarz have shown tight
bounds for solving approximate and asymptotic Consensus in quite
general message adversaries. 

In \cite{consensus-epistemo}, Nowak, Schmid, and Winkler propose knowledge-based
topologies for all message adversaries. It is then used to characterize
message adversaries that can solve Consensus. The scope of
\cite{consensus-epistemo} is larger than the scope of the paper, however, note that contrary to
those topologies, that are \emph{implicitly} defined by indistinguishability
of local knowledge, the geometrization topology here is explicitly defined
and fully described by Th.~\ref{thm:geoEquiv}. Moreover, set-agreement has not
been investigated in this knowledge-based framework.

The $k-$set agreement problem is a classical and important
coordination problems. It is also a theoretical benchmark for
distributed computability in numerous models.  A review by Raynal can
be found in \cite{ksetRaynal}. Set-agreement is the weaker version of
$k-$set agreement, \ie with $k+1$ processes only.

In \cite{2generals-journal}, the ``two generals problem'', that is the
Consensus problem for two processes is investigated for arbitrary
sub-IIS models by Godard and Perdereau. Given that Consensus for two
processes is actually the set-agreement problem, the characterization of
solvability of set-agreement presented here is a generalization to
any number of processes of the results of
\cite{2generals-journal}.

\subsection{Outline}

In Section~\ref{sec:std-defs}, we present standard definitions and our
notation for message adversaries. In Section~\ref{sec:topo}, we start
by restating the standard definitions of using combinatorial topology
for distributed computing in the geometric context. We then introduce
the geometrization topology of $IIS_n$. In Section~\ref{sec:geoequiv},
we describe and prove precisely the properties of the geometrization
topology.  In Section~\ref{sec:setag}, we apply our framework to
derive new characterization of computability and lower bounds for set-agreement.
We conclude in Section~\ref{sec:ccl} with some open questions.

\section{Models and Definitions}
\label{sec:std-defs}

\subsection{Message Adversaries}
\label{subsec:def}
We introduce and present here our notations. Let $n\in\N$, we consider
systems with $n+1$ processes. We denote $\Pi_n=[0,..,n]$ the set of
processes. Since sending a message is an asymmetric operation, we
will work with directed graphs. We recall the main standard
definitions in the following.

We use standard directed graph (or digraph) notations : given $G$, $V(G)$ is
the set of vertices, $A(G)\subset V(G)\times V(G)$ is the set of arcs.

\begin{definition}
  We denote by $\mathcal G_n$  the set of directed graphs with vertices
  in $\Pi_n$.

  A \emph{dynamic graph} \dG is a sequence $G_1,G_2,\cdots,G_r,\cdots$ where
  $G_r$ is a directed graph with vertices in $\Pi_n$. We also denote by $\dG(r)$ the digraph $G_r$.
  A \emph{message adversary} is a set of dynamic graphs.
\end{definition}

Since that $n$ will be mostly fixed through the paper, we use $\Pi$
for the set of processes and $\mathcal G$ for the set of graphs with
vertices $\Pi$ when there is no ambiguity.  

\medskip
Intuitively, the graph at position $r$ of the sequence describes
whether there will be, or not, transmission of some messages sent at
round $r$. A formal definition of an execution under a scenario will
be given in Section~\ref{execution}.

We will use the standard following notations in order to describe more easily
our message adversaries \cite{PPinfinite}. A sequence is
seen as a word over the alphabet $\mathcal G$.
The empty word is noted $\varepsilon$.

\begin{definition}
  Given $A\subset\mathcal G$, $A^*$ is the set of all finite
  sequences of elements of $A$, $A^\omega$ is the set
  of all infinite ones and $A^\infty = A^* \cup A^\omega.$
\end{definition}

Given $\dG\in\mathcal G^\omega$, if $\dG=\dH\dK$, with $\dH\in\mathcal G_n^*,
\dK\in\mathcal G_n^\omega$, we say that $\dH$ is \emph{a prefix} of $\dG$, and $\dK$
\emph{a
suffix}. $Pref(\dG)$ denotes the set of prefixes of $\dG$.  An adversary of
the form $A^\omega$ is called an \emph{oblivious adversary} or an
\emph{iterated adversary}.  A word in $\ma\subset\mathcal G^\omega$ is
called a \emph{communication scenario} (or \emph{scenario} for short)
of message adversary $\ma$.  Given a word $\dH\in\mathcal G^*$, it is
called a \emph{partial scenario} and $len(\dH)$ is the length of this
word. The prefix of $\dG$ of length $r$ is denoted $\dG_{\mid r}$ (not
to be confused with $\dG(r)$ which is the $r$-th letter of $\dG$, it the digraph at time $r$).

The following definitions provide a notion of causality when considering infinite word over digraphs.

\begin{definition}[\cite{tvg}]\label{def:causal}
  Let \dG a sequence $G_1,G_2,\cdots,G_r,\cdots$. Let
  $p,q\in\Pi$. There is a \emph{journey} in \dG at time $r$ from $p$ to $q$,
  if there exists a sequence $p_0,p_1,\dots,p_t\in\Pi$, and a sequence
  $r\leq i_0<i_1<\cdots<i_t\in\N$ where we have
  \begin{itemize}
  \item $p_0=p,p_t=q$,
  \item for each $0 < j \leq t$, $(p_{j-1},p_j)\in A(G_{i_j})$
  \end{itemize}
  This is denoted $p\causal\limits^r_\dG q$. We also say that $p$ is
  causally influencing $q$ from round $r$ in \dG.
\end{definition}

\subsection{Iterated Immediate Snapshot Message Adversary}
\label{sect:iis}
We say that a graph $G$ has the \emph{Immediacy Property} if for all $a,b,c\in V(G)$, $(a,b),
(b,c)\in A(G)$ implies that $(a,c)\in A(G)$.
A graph $G$ has the \emph{containment Property} if for all $a,b\in
V(G)$, $(a,b)\in A(G)$ or $(b,a)\in A(G)$.

\begin{definition}[\cite{messadv,HKRbook}]
  We set
  $IS_n = \{G\in\mathcal G_n\mid G \mbox{ has the Immediacy and
 } \mbox{Con\-tainment properties}\}.$

The Iterated Immediate Snapshot message adversary for $n+1$ processes is the
message adversary $IIS_n=IS_n^\omega$.
\end{definition}

The Iterated Immediate Snapshot model was first introduced as a
(shared) memory model and then has been shown to be equivalent to the
above message adversary first as tournaments and iterated tournaments \cite{BGequivIIS,messadv}, then
as this message adversary \cite{HKRbook,DCcolumn}.  See also
\cite{Rajsbaum-iterated} for a survey of the reductions involved in these layered
models.

\subsection{Examples}
\label{examples}
  We do not restrict our study to regular languages, however all message
  adversaries we are aware of are regular, as can be seen in the
  following examples, where the regular expressions prove to be very
  convenient.
We show how standard fault environments are conveniently described in
our framework.%

\begin{example}
  Consider a message passing system with $n$ processes where, at each round, all messages can be
  lost. The associated message adversary is $\mathcal G_{n-1}^\omega$.
\end{example}

\begin{example}
  Consider a system with two processes $\{\b,\n\}$ where, at each round, only one message can be
  lost. The associated message adversary is
  $\{\lok,\lblanc,\lnoir\}^\omega$. This is $IIS_1$.
\end{example}

\subsection{Execution of a Distributed Algorithm}
\label{execution}
Given a message adversary $\ma$ and a set of initial configurations \I,
we define what is an execution of a given algorithm \algo subject to $\ma$ with initialization \I.
An execution is constituted of an initialization step, and a (possibly infinite) sequence of rounds of
messages exchanges and corresponding local state updates.
When the initialization is clear from the context, we will use
\emph{scenario} and \emph{execution} interchangeably.

An execution of an algorithm \algo under scenario $w\in \ma$ and initialization $\iota\in\I$
is the following. This execution is denoted $\iota.w$.
First, $\iota$ affects an initial state to all
processes of $\Pi$.

A round is decomposed in 3 steps : sending, receiving, updating the
local state.  At round $r\in\N$, messages are sent by the processes
using the \texttt{SendAll()} primitive. The fact that the corresponding
receive actions, using the \texttt{Receive()} primitive, will be successful depends on $G=w(r)$,
$G$ is called the \emph{instant graph} at round $r$.

Let $p,q\in\Pi$. The message sent by $p$ to all its neighbours is
received by $q$ on the condition that the arc $(p,q)\in A(G)$.
Then, all processes update their state according to the
received values  and \algo. 
Note that, it is usually assumed that $p$ always receives its own
value, that is $(p,p)\in A(G)$ for all $p$ and $G$. 

\medskip
Let $w\in\ma, \iota\in\I$. Given $u\in Pref(w)$, we denote by $\mathbf s_p(\iota.u)$ the
state of process $p$ at the $len(u)$-th round of the algorithm \algo
under scenario $w$ with initialization $\iota$. This means in particular that
$\mathbf s_p(\iota.\varepsilon)$ represents the initial state of $p$ in $\iota$, where
$\varepsilon$ denotes the empty word.

A task is given by a set \I of initial configurations, a set of output
values $Out$ and a relation $\Delta$, the specification, between
initial configurations and output configuration\footnote{Note that the
  standard definition in the topological setting involves carrier map
  that we do not consider here for we will consider only one specific
  task, the Set Agreement problem}.  We say that a process
\emph{decides} when it outputs a value in \O.  Finally and
classically,
\begin{definition}
An algorithm \algo solves a Task $(\I,Out,\Delta)$ for the message
adversary $\ma$ if for any $\iota\in\I$, any scenario $w\in \ma$,
there exist $u$ a prefix of $w$ such that the states of the processes
$out = (\mathbf s_0(\iota.u),\dots,\mathbf s_n(\iota.u))$ satisfy the
specification of the task, ie $\iota \Delta out$.
\end{definition}

\section{A Topology by Geometrization}
\label{sec:topo}

In this paper we present a new topological approach for investigating
distributed computability.  It extends the known simplicial
complexes-based known method for finite executions to infinite
executions without considering infinite additional complexes like in
\cite{GKM14}.  This enables to define directly a topology on the set of
executions of the standard Iterated Immediate Snapshot model $IIS_n$.

\subsection{Combinatorial Topology Definitions}

\subsubsection{Geometric Simplicial Complexes}

Before giving the definition of the geometrization topology in
Sect.~\ref{sect:geo}, we state the definition of simplicial complexes,
but not first as abstract complex, as is usually done in distributed
computing, but primarily as geometrical objects in $\R^N$. This is the
reason we call this definition the geometrization topology.
Intuitively we will associate%
a point in $\R^N$ to any execution via a geometrization mapping $geo$.
The geometrization topology is the topology induced by $geo^{-1}$ from
the standard topology in $\R^N$. This also makes $geo$ continuous by
definition.  In the standard approach, geometric simplices are also
used but they are introduced as geometric realizations of the
abstract simplicial complexes.  As will be seen later, when dealing
with infinite complexes, the standard topology of these simplices does
not enable to handle the computability of distributed tasks since we
will need to define an other topology. 
We show that the topology on infinite complexes, as defined in
standard topology textbook, is different from the one we show here to
be relevant for distributed computability. See Section~\ref{sec:contrex}.
Note that to be correctly
interpreted, the topology we construct is on the set of infinite
executions, not on the complexes corresponding the finite executions.

The following definitions are standard definitions from algebraic topology \cite{Munkres84}.
We fix an integer $N\in\N$ for this part. 
 We denote $||x||$ the euclidean
norm in $\R^N$. For a bounded subset $X\subset\R^n$, we denote $diam(X)$ its diameter.

\begin{definition}
  Let $n\in\N$.
  A finite set $\sigma=\{x_0,\dots,x_n\}\subset\R^N$ is called a \emph{simplex} of dimension $n$ if the vectors
  $\{x_1-x_0,\dots,x_n-x_0\}$ are linearly independent.
 We denote by $|\sigma|$ the convex hull of $\sigma$.
\end{definition}

\begin{definition}[\cite{Munkres84}]
  \label{def:simcomplex}
A \emph{simplicial complex} is a collection $C$ of \emph{simplices}
such that~:
\begin{enumerate}[(a)]
  \item If $\sigma\in C$ and $\sigma'\subseteq\sigma$, then $\sigma'\in C$,
  \item If $\sigma,\tau\in C$ and $|\sigma|\cap|\tau|\neq\emptyset$ then there exists $\sigma'\in C$ such that
    $|\sigma|\cap|\tau|=|\sigma'|$,
    $\sigma'\subset\sigma, \sigma'\subset\tau.$
\end{enumerate}
\end{definition}

We denote $\wr C\wr=\bigcup\limits_{S\in C}|S|$, this is the
\emph{geometrization of $C$}.

Note that these definitions do not require complexes to be a finite collection of simplicies.
The simplices of dimension 0 (singleton) of $C$ are called vertices,
we denote $V(C)$ the set of vertices of $C$.
A complex is pure of dimension $n$ if all maximal simplices are of dimension $n$.
In this case, a simplex of dimension $n-1$ is called a facet.
The \emph{boundary} of a simplex $\sigma=\{x_0,\dots,x_n\}$ is the pure
complex $\bigcup_{i\in[0,n]}\{x_j\mid j\in[0,n], i\neq j\}$ of
dimension $n-1$. It is denoted $\delta(\sigma)$, it is the union of
the facets of $\sigma$.

Let $A$ and $B$ be simplicial complexes. A map $f\colon V(A)\to V(B)$
defines a  \emph{simplicial map} 
if it preserves the simplices, \ie for each simplex $\sigma$ of $A$,
the image $f(\sigma)$ is a simplex of $B$. By linear combination of the barycentric coordinates, $f$ extends to
the linear simplicial map $f\colon \wr A\wr\to \wr B\wr$, which is continuous. See \cite[Lemma 2.7]{Munkres84}.

We also have colored simplicial complexes. These
are simplicial complexes $C$ together with 
a function $\chi:V(C)\to \Pi$ such that the restriction of $\chi$ on any
maximal simplex of $C$ is a bijection. A simplicial map that preserves colors is called chromatic.

Finally, $S^n$ will denote ``the'' simplex of dimension $n$. Through
this paper we assume a fixed embedding in $\R^N$ for $S^n=(x_0^*,\dots,x_n^*)$.
We will also assume that its diameter $diam(S^n)$ is 1.

\subsubsection{The Standard Chromatic Subdivision} \label{sect:chr}

Here we present the standard chromatic subdivision, \cite{HKRbook} and
\cite{Kozlovbook}, as a geometric complex. We start with subdivisions and chromatic subdivisions.

\begin{definition}[Subdivision]
  A subdivision of a simplex
  $S$ is a simplicial complex $C$ with $\wr C\wr=|S|$.
\end{definition}

\begin{definition}[Chromatic Subdivision]
  Given $(S,\mathcal P)$ a chromatic simplex, a chromatic subdivision of
  $S$ is a chromatic simplicial complex $(C,\mathcal P_C)$ such that
\begin{itemize}
\item $C$ is a subdivision of $S (\ie $$\wr C\wr=|S|$),
\item $\forall x\in V(S), \mathcal P_C(x)=\mathcal P(x).$
\end{itemize}
\end{definition}

Note that it is not necessary to assume $V(S)\subset V(C)$ here, since
the vertices of the simplex $S$ being extremal points, they are
necesarily in $V(C)$.

We start by defining some geometric transformations of simplices (here
seen as sets of points). The choice of the coefficients will be justified later.
\begin{definition}\label{def:geo}
  Consider a set $V=(y_0,\dots,y_d)$ of size $d+1$ in $\R^N$. We
  define the function $\zeta_V:V\longrightarrow
  \mathcal \R^N$ by, for all $j\in[0,d]$
 $$ \zeta_V(y_j) = \frac{1-\frac{d}{2d+1}}{d+1}y_j + \sum_{i\neq j}\frac{1+\frac{1}{2d+1}}{d+1}y_i $$

\end{definition}

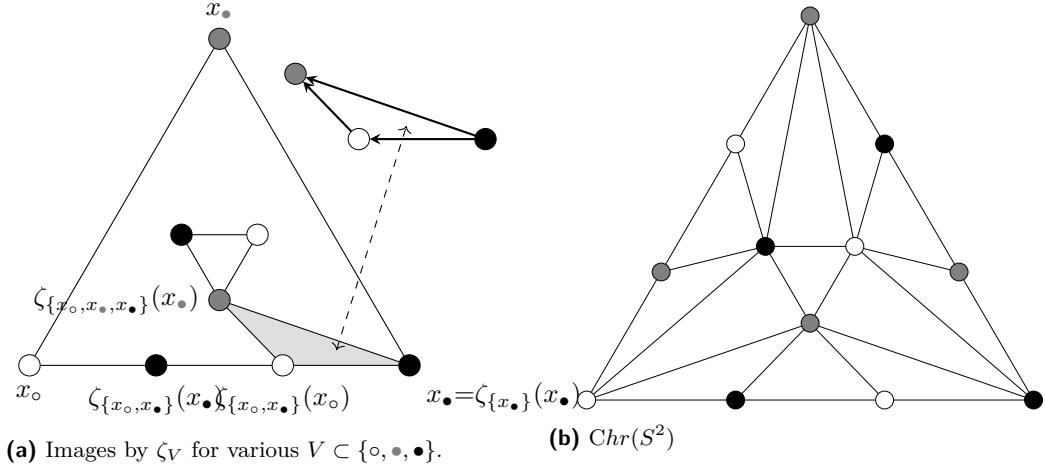
\begin{figure}[t] %
  \centering
  \begin{subfigure}{.49\textwidth}
  \centering
  \begin{tikzpicture}[scale=.5]
      \path[fill=lightgray,opacity=0.5] (10,0) -- (5.0, 1.7320508075688772) -- (6.6666666,0);

    \node[proc,g,label=north:$x_\g$] (0) at (5.0, 8.660254037844386) {};
      \node[proc,b,label=south:$x_\b$] (1) at (0, 0) {};
      \node[proc,n,label=south east:$x_\n {=} \zeta_{\{x_\n\}}(x_\n)$] (2) at (10, 0) {};
      \node[proc,g,label=left:$\zeta_{\{x_\b,x_\g,x_\n\}}(x_\g)$] (3) at (5.0, 1.7320508075688772) {};
      \node[proc,b] (4) at (6.0, 3.4641016151377544) {};
      \node[proc,n] (5) at (4.0, 3.4641016151377544) {};
      \node[proc,n,label=south:$\zeta_{\{x_\b,x_\n\}}(x_\n)$] (6) at (3.3333333,0) {};
      \node[proc,b,label=south:$\zeta_{\{x_\b,x_\n\}}(x_\b)$] (7) at
      (6.6666666,0) {};
      \node (11) at (8,0.15) {};
      \draw (3) -- (4); %
      \draw (4) -- (5); %
      \draw (5) -- (3); %
      \draw (1) -- (6);
      \draw (6) -- (7); %
      \draw (7) -- (2);
      \draw (0) -- (2);
      \draw (1) -- (0);
      \draw (3) -- (7);
      \draw (3) -- (2);

      \node[proc,b] (8) at (8.66666666,6) {};
      \node[proc,n] (9) at (12,6) {};
      \node[proc,g] (10) at (7,7.7320508075688772) {};
      \node (12) at (10,6.6) {};
      \draw[rel,thick,->] (9) -- (8);
      \draw[rel,thick,->] (9) -- (10);
      \draw[rel,thick,->] (8) -- (10);
      \draw[dashed,<->] (11) -- (12);
  \end{tikzpicture}
  \caption{Images by $\zeta_V$ for various $V\subset\{\b,\g,\n\}$.}
  \label{fig:chr_zeta}
  \end{subfigure}%
  \hfill
  \begin{subfigure}{.49\textwidth}
  \centering
    \includegraphics[scale=1.05]{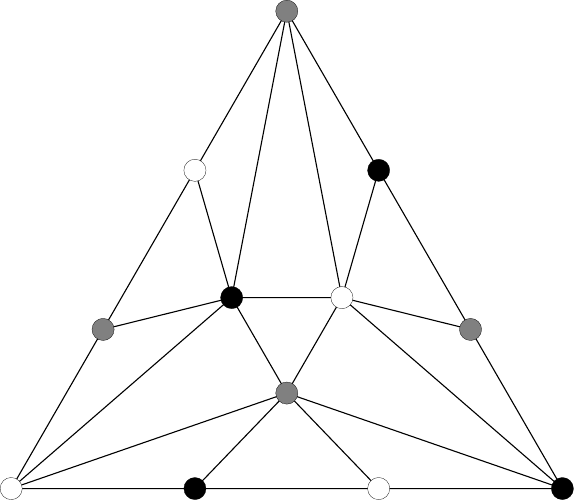}
    \caption{$\chr(S^2)$}
    \label{fig:sub_chr_3d}
  \end{subfigure}
  \caption{Standard chromatic subdivision construction for dimension
    $2$. On the left, the association between an instant graph of
    $IS_2$ (top) and a simplex of $\chr(S^2)$ (grey area) is illustrated.}
  \label{fig:sub_chr_2et3d}
\end{figure}

We now define in a geometric way the \emph{standard chromatic
  subdivision} of colored simplex $(S,\mathcal P)$, where
$S=\{x_0,x_1,\dots,x_n\}$ and $\mathcal P(x_i)=i$. 

The chromatic subdivision $\chr(S)$ for the colored simplex
$S=\{ x_0,\dots,x_n \}$ is a simplicial complex defined by the set of
vertices
$V(\chr(S)) = \{\zeta_{V}(x_i) \mid  i\in [0,n], V\subset V(S), x_i\in V\}.$

For each pair $(i,V)$, $i\in [0,n]$ and $V\subset V(S)$,
there is an associated vertex $y$ of $\chr(S)$, and conversely each
vertex has an associated pair. The \emph{color} of $(i,V)$ is
$i$. The set $V$ is called the \emph{view}.
We define $\Phi$ the following \emph{presentation} of a vertex $y$,
$\Phi(y)=(\mathcal P(y),V_y)$ where $\mathcal P(y) = i$ and $V_y=V$.

The simplices of $\chr(S)$  are the set of $d+1$ points $\{\zeta_{V_0}(x_{i_0}),\cdots,\zeta_{V_d}(x_{i_d})\}$ where
\begin{itemize}
\item there exists a permutation $\pi$ on $[0,d]$ such that
  $V_{\pi(0)}\subseteq\cdots\subseteq V_{\pi(d)}$,
\item If ${i_j}\in\mathcal P(V_\ell)$ then $V_j\subset{V_\ell}$.
\end{itemize}

In Fig.~\ref{fig:sub_chr_2et3d}, we present the construction for
$\chr(S^2)$. For convenience, we associate $\b,\g,\n$ to the processes $0,1,2$
respectively. In Fig.~\ref{fig:chr_zeta}{}, we consider the triangle $x_\b, x_\g, x_\n$ in $\R^2$,
with $x_\b=(0,0)$, $x_\n=(1,0),$ $x_\g=(\frac{1}{2},\frac{\sqrt{3}}{2})$. We have that
$\zeta_{\{x_\b,x_\n\}}(x_\n)=(\frac{1}{3},0)$,
$\zeta_{\{x_\b,x_\n\}}(x_\b)=(\frac{2}{3},0)$
and $\zeta_{\{x_\b,x_\g,x_\n\}}(x_\g) = (\frac{1}{2},\frac{\sqrt{3}}{5})$.
The relation between instant graph (top) and simplex
$\left\{(\frac{2}{3},0),(1,0),(\frac{1}{2},\frac{\sqrt{3}}{5})\right\}$ (grey area) is
detailed in the following section.

In the following, we will be interested in iterations of
$Chr(S^n,\mathcal P)$. The last property of the definition of
chromatic subdivision means with we can drop the $C$ index in the
coloring of complex $C$ and use $\mathcal P$ to denote the coloring at
all steps. From its
special role, it is called the \emph{process color} and we drop
$\mathcal P$ in $Chr(S,\mathcal P)$ using in the following $Chr(S)$
for all simplices $S$ of iterations of $Chr(S^n)$.

In \cite{Koz12}, Kozlov showed how the standard chromatic subdivision complex relates to Schlegel
diagrams (special projections of cross-polytopes), and used this relation
to prove the standard chromatic subdivision was actually a
subdivision.

In \cite[section 3.6.3]{HKRbook}, a general embedding in $R^{n}$
parameterized by $\epsilon\in\R$ is given for the standard chromatic
subdivision. The geometrization here is done choosing $\epsilon =
\frac{d}{2d+1}$ in order to have ``well balanced'' drawings.

\subsection{Encoding Iterated Immediate Snapshots Configurations}

\subsubsection{Algorithms in the Iterated Immediate Snapshots Model}
It is well known, see \eg \cite[Chap. 3\&4, Def. 3.6.3]{HKRbook}, that each maximal simplex
$S=\{\zeta_{V_0}(x_{i_0}),\cdots,\zeta_{V_n}(x_{i_n})\}$
from the chromatic subdivision of $S^n$ can be associated with
a graph of $IS_{n}$ denoted $\Theta(S)$.
We have $V(\Theta(S))=\Pi_n=[0,n]$ and set
$\Theta(\zeta_{V_j}(x_{i_j}))=\mathcal P(x_{i_j})$.%
The arcs are defined using the representation $\Phi$ of points,
$A(\Theta(S))= \{(i,j)\mid i\neq j, V_i\subseteq V_j\}.$
The mapping $\theta$ will denote $\Theta^{-1}$.
We can transpose this presentation to an averaging algorithm called
the \emph{Chromatic Average} Algorithm presented in Algorithm~\ref{alg:std-loop}.
\SetKwFor{Loop}{Loop}{}{EndLoop}

\begin{algorithm}%
  $x \leftarrow x_i^*$\;
  \Loop{forever}{
    \texttt{SendAll}$((i,x))$\;
    $V\leftarrow$\texttt{Receive()} // set of all received messages\; %
    $d\leftarrow sizeof(V) - 1$ // $i$ received $d+1$ messages
    including its own \; 
    $x = \frac{1-\frac{d}{2d+1}}{d+1}x + \sum_{(j,x_j)\in V, j\neq i}\frac{1+\frac{1}{2d+1}}{d+1}x_j$\;
  }
  \caption{The Chromatic Average Algorithm for process $i$\label{alg:std-loop}}
\end{algorithm}

Executing one round of the loop in
Chromatic Average for instant graph $G$, the state of
process $i$ is $x'_i=\zeta_{V_i}(x^*_{i})$, where $V_i$ is the
view of $i$ on this round, that is the set of $(j,x_j)$ it has received; with 
$\Theta(\{\zeta_{V_0}(x^*_{0}),\cdots,\zeta_{V_n}(x^*_{n})\})=G$.
See eg. in Fig.~\ref{fig:chr_zeta}, the simplex of the grey area
corresponds to the ordered sequence of views $\{x_\n\}\subset\{x_\n,x_\b\}\subset\{x_\n,x_\b,x_\g\}$,
associated to the directed graph depicted at the top right.
Adjacency for a given $i$ corresponds to the smallest subset
containing $x_i$. 
By iterating, the chromatic subdivisions $\chr^r(S^n)$
are given by the global state under all possible $r$ rounds of the Chromatic
Average Algorithm.
Finite rounds give the Iterated Chromatic Subdivision (hence the name). This is an algorithm that is not meant to terminate (like the full information protocol). The infinite runs are used below to define a
topology on $IIS_n$.

The Chromatic Average algorithm is therefore the geometric counterpart
to the Full Information Protocol that is associated with $\chr$
\cite{HKRbook}. In particular, any algorithm can be presented as the
Chromatic Average together with a terminating condition and an output
function of $x$.

This one round transformation for the canonical $S^n$ can actually be
done for any simplex $S$ of dimension $n$ of $\R^N$.
For $G \in IS_n$, we denote
$\mu_G(S)$ the geometric simplex that is the image of $S$ by one round of the Chromatic average
algorithm under instant graph $G$.

\medskip
The definitions of the previous
section can be considered as mostly textbook (as in \cite{HKRbook}), or
folklore. %
To the best of our knowledge, the Chromatic Average Algorithm, as such, is new,
and there is no previous complete proof of the link between the Chromatic Average Algorithm and
iterated standard chromatic subdivisions.
However, one shall remark that people are, usually, actually \emph{drawing}
standard chromatic subdivisions using the Chromatic Average Algorithm.

\subsubsection{A Topology for $IIS_n$}
\label{sect:geo}

Let $w\in IIS_n$, $w=G_1G_2\cdots$. For the prefix of $w$ of size $r$,
$S$ a simplex of dimension $n$, we define $geo(w_{\mid r})(S) =
\mu_{G_r}\circ\mu_{G_{r-1}}\circ\cdots\circ\mu_{G_1}(S)$.
Finally, we set
$geo(w) = \mathop{\lim}\limits_{r\longrightarrow\infty}geo(w_{\mid r})(S^n)$.  We prove
in the following section that this actually converges.

We define the \emph{geometrization topology} on the space $IIS_n$ by
considering as open sets the sets $geo^{-1}(\Omega)$ where $\Omega$ is
an open set of $\R^N$.  A collection of sets can define a topology
when any union of sets of the collection is in the collection, and
when any finite intersection of sets of the collection is in the
collection. This is straightforward for a collection of inverse images
of a collection that satisfies these properties.

\medskip
 A neighbourhood for point $x$ is an open set containing $x$.
 In topological spaces, a pair of distinct points $x,y$ is called
 \emph{non-separable} if there does not exist two disjoint
 neighbourhoods of these points.  The pre-images
 $geo^{-1}(x)$ that are not singletons are \emph{non-separable
   sets}. We will see that we always have non-separable sets and that
 they play an important role for task solvability.

Subset of $IIS_n$ will get the subset topology, that is , for $\ma\subseteq IIS_n$,
open sets are the sets $geo^{-1}(\Omega)\cap\ma$ where $\Omega$
is an open set of $\R^N$.
We set $\wr\ma\wr = geo(\ma)$ the geometrization of $\ma$. 

Note that the geometrization should not be confused with the standard
\emph{geometric realization}. They are the same at the set level but
not at the topological level. This is quite well known, see e.g. \cite{Kozlovbook}.
At times, in order to emphasize this difference, for a simplex $S\subset\R^N$,
we will also use $\wr S\wr$ instead of $|S|$.  The
\emph{geometrization} of $C$, denoted $\bigwr C\bigwr$, that is the union
of the convex hulls $|\sigma|$ of the simplices $\sigma$ of $C$,
is endowed with the standard topology from $\R^N$.
We also note this topological space as $\wr C\wr$.

\begin{remark}
  When considering the input complex embedded in $\R^N$, the
  geometrization topology could be applied on all simplices, in effect
  providing a new topological framework for so called protocol
  complex. This could be done by applying the Chromatic Average
  algorithm.  This is not detailed here as we will not need it to
  investigate set-agreement. Moreover, note that this construction
  works also for any model that corresponds to a mesh-shrinking subdivision.
\end{remark}

\subsubsection{Convexity and Metric Results}
\label{sec:converge}
Before investigating the geometrization topology, we present some
metric results relating vertices of the chromatic subdivision. In
particular we prove that the sequences $geo(w_{\mid r})(S)$ converge
to a point.

  The following lemma comes from the convexity of the $\mu_G$ transforms.
\begin{lemma}\label{lemma:convexSimplex}
Let $w$ a run, let $r,r'\in\N, r<r'$ then $| geo(w_{\mid r'})(S^n)|\subset | geo(w_{\mid r})(S^n)|.$
\end{lemma}
\begin{proof}
Consider only one step. We have that $\frac{1-\frac{d}{2d+1}}{d+1} +
d\times\frac{1+\frac{1}{2d+1}}{d+1} =
\frac{1-\frac{d}{2d+1}+d+\frac{d}{2d+1}}{d+1}=1$.  So one step of the
Chromatic Average gives, on each process, a linear combination with
non-negative coefficients that sums to 1, it is therefore a
barycentric combination on the points of the simplex at the beginning
of the round. It is therefore a convex mapping of this simplex. Since
composing convex mapping is also convex, and that $S^n$ is a convex
set, we get the result by recurrence.
\end{proof}

\begin{lemma}\label{lem:metricineq}
  There exists reals $0<K'<K<1$, such that for all $G$ of
  $IS_n$, all $p,q\in V(S)$, $p',q'\in
  V(\mu_G(S))$, such that $\mathcal P(p)=\mathcal P(p')$ and $\mathcal
  P(q)=\mathcal P(q')$, we have
  $$K'||p-q||\leq ||p'-q'||\leq K||p-q||.$$
\end{lemma}
\begin{proof}
  This is a consequence of $\mu_G$ transforms being
  convex when $G\in IS_n$.  It corresponds to a stochastic matrix
  (non-negative coefficients and all lines coefficient sums to one)
  that is scrambling (there is a line without null coefficients) hence
  contractive. See \eg \cite[Chap. 1]{Hartfiel98} for definitions and a proof
  for any given $G$ of $IS_n$.

  Then $K$ (resp. $K'$) is the largest (resp. smallest) such bounds over all $G\in IS_n$. 
\end{proof}

While iterating the chromatic subdivision, we remark that the diameter
of the corresponding simplices is contracting. From Lemma~\ref{lem:metricineq}, we have
\begin{lemma}\label{lem:contraction}
Let $S$ a simplex of $\R^\N$, then
$diam(\mu_{G_r}\circ\mu_{G_{r-1}}\circ\cdots\circ\mu_{G_1}(S))
\leq {K^r} diam(S)$, where $K$ is the constant from the previous lemma.
\end{lemma}

Since the simplices are contracted by the $\mu_G$ functions, the
sequence of isobarycenters of \\
$(geo(w_{\mid r}(S))_{r\in\N^*}$ has
the Cauchy property and this sequence is therefore convergent to some
point $x\in\R^N$. Since the diameter of the simplices converges to 0, it makes senses
to say that the limit of the simplices is the point $x$. Note that it
would also be possible to formally define a metric on the convex
subsets of $\R^N$ and consider the convergence of the simplices in
this space.

\section{Geometrization Equivalence}
\label{sec:geoequiv}
As will be be shown later, the geometrization has a crucial role in
order to understand the relationship between sets of possible
executions and solvability of distributed tasks.  In this section, we
describe more precisely the pre-images sets, that is subsets of $IIS_n$
of the form $geo^{-1}(x)$ for $x\in|S^n|$.
In particular, we will get a description of the non-separable sets of execution.

\subsection{Definitions}

We say that two executions $w,w'\in IIS_n$ are
\emph{geo-equivalent} if $geo(w)=geo(w')$.  The set of all
$w'$ such that $geo(w)=geo(w')$ is called the equivalence class of
$w$.  Since the topology we are interested in for $\wr  IIS_n\wr$ is the
one induced by the standard separable space $\R^N$ via the $geo^{-1}$ mapping, it is
straightforward to see that non-separable sets are exactly the
geo-equivalence classes that are not singletons.
In this section, we describe all equivalence classes and show that
there is a finite number of possible size for these sets.

\medskip
We define subsets of $IS_n$, the sets $Solo(P)$, that correspond to
instant graphs where the process in $P\subset\Pi$ have no incoming
message from processes outside of $P$. We have $Solo(\Pi)=IS_n$.

\begin{definition}\label{def:Solo}
  In the complex $\chr(S^n)$, with $P\subset\Pi$, $Solo(P)$ is the set of 
  simplices $T\in\chr(S^n)$ such that $\forall (p,q)\in A(\Theta(T)), q\in P\Rightarrow p\in P$.
\end{definition}

We denote by $\mathcal K_\Pi$ the instant graph that is complete on $\Pi$.

  An execution $w$ is said \emph{fair} for $P$, $w\in Fair(P)$, if
  $w\in Solo(P)^\omega$ and for all $p,q\in P$, $\forall r\in \N$, we
  have $p\causal\limits^r_w q$. Fairness for $P$ means that processes
  in $P$ are only influenced by processes in $P$, and that any process
  always influence other processes infnitely many times.
We have the equivalent, and constructive definition :

\begin{proposition}\label{prop:FairTopo}
  Let $w\in Solo(P)^\omega$. An execution $w$ is $Fair$ for $P$ if
  and only if $w$ has no suffix in $\bigcup_{Q\neq\emptyset, Q\subsetneq
    P}Solo(Q)^\omega$.
\end{proposition}

\begin{proof} 
  Assume we have a suffix $s$ for $w$ in $Solo(Q)^\omega$ with
  ${Q\subsetneq P}$.Let $p\in P\setminus Q$ and $r$ the starting index
  of the suffix.  Then $\forall q\in Q$, we must have $p\causal\limits^r_w q$
  by definition of fairness for $P$.  Denote $q_0$ the first element of $Q$ to be
  causally influenced by $p$ at some time $t\geq r$.  So $q_0$ receive a
  message from some $p'\in\Pi$, $p'\neq q_0$ at time $t$.  Since $s\in
  Solo(Q)^\omega$, this means that $q_0$ can only receive message from
  processes in $Q$. Hence $p'\in Q$ and $p'$ was influenced by $p$ at
  time $t-1$. A contradiction with the minimality of $q_0$.  So $w$ is
  not in $Fair(P)$.

  Conversely, assume that $w\notin Fair(P)$. Then $\exists p,q\in
  P,\exists s, \forall r\geq s$, $\neg p\causal\limits^r_w q$. We set $Q$ as
  the set of processes that causally influence $q$ for all $r\geq
  s$. We have $p\notin Q$ so $Q\subsetneq P$.  We denote $s_0$, a time
  at which no process of $\Pi\setminus Q$ influence a process in
  $Q$. By construction, the suffix at step $s_0$ is in
  $Solo(Q)^\omega$.
\end{proof}

\subsection{First Results on Geometrization}

We start by presenting a series of results about geometrization.
Lemma~\ref{lemma:convexSimplex} gives the following immediate corollaries. 
\begin{corollary}\label{prop:geoinsimplex}
  Let $w$ a run in $IIS_n$, then $\forall r \in \N,$ $geo(w) \in | geo(w_{\mid r})(S^n)|.$
\end{corollary}

\begin{proposition}\label{prop:geoEquivalenceIntersection}
Let $w,w'$ two geo-equivalent runs in $IIS_n$, then $\forall r \in
\N,$ $geo(w_{\mid r})(S^n)\;\cap\;geo(w'_{\mid r})(S^n)\neq\emptyset$.
\end{proposition}
\begin{proof}
  The intersection of the geometrizations $| geo(w_{\mid
    r})(S^n)|$ and $| geo(w'_{\mid r})(S^n)|$ contains at least
  $geo(w)$ by previous corollary. Since the simplices $geo(w_{\mid
    r})(S^n)$ and $geo(w'_{\mid r})(S^n)$ belong to the complex
  $\chr^r(S^n)$, they also intersect as simplices.
\end{proof}

\begin{proposition}
  Let $S$ a maximal simplex of the chromatic subdivision $\chr S^n$ that is not $\theta(\mathcal K(\Pi))$.
  Then there is $P\subsetneq\Pi$ such that $\Theta(S)\in Solo(P)$.
\end{proposition}

Conversely we can describe $Solo(P)$ more precisely. We denote by
$\delta(S^n,P)$ the sub-simplex of $S^n$ corresponding to $P\subset\Pi$. This is the boundary relative to $P$ in $S^n$, and we have that $\bigcup_{P\subsetneq\Pi}\delta(S^n,P) = \bigcup_{p\in\Pi}\delta(S^n,\pi\setminus{p}) = \delta(S^n)$.
\begin{proposition}[Boundaries of $\chr$ are $Solo$]\label{prop:soloboundary}
  Let $P$ a subset of $\Pi.$
  Then $Solo(P)=\{S\mid S\mbox{ a maximal simplex of }\chr(S^n) , |S|\cap|\delta(S^n,P)|\neq\emptyset\}$.
\end{proposition}
\begin{proof}
  Denote $q$ such that $\Pi=P\cup\{q\}$. Then by construction,
  $Solo(P)$ corresponds exactly to the simplex where the processes in
  $P$ do not receive any message from $q$, ie the simplex intersecting
  the boundary $\delta(S^n,P)$.
\end{proof}
For a given size $s$ of $P$, the $Solo$ sets are disjoint, however 
this does not form a partition of $\chr$.
Finally, by iterating the previous proposition, the
boundaries of $S^n$ are described by $Solo(P)^\omega$.
\begin{proposition}\label{prop:solomegaboundary}
  Let $P$ a subset of $\Pi$. We have 
 $\wr Solo(P)^\omega\wr = |\delta(S^n,P)|.$
\end{proposition}

We can now state the main result that links geometrically fair
executions and corresponding simplices : in a fair execution, the
corresponding simplices, that are included by convexity, have to
eventually be strictly included in the interior.

\begin{proposition}[Geometric interpretation of $Fair$]\label{prop:geoFair}
Let $w$ an execution that is $Fair$ for $\Pi$, then $\forall
s\in\N,\exists r>s \in \mathbb{N}$, such that $\delta(geo(w_{\mid s})(S))
\cap geo(w_{\mid r})(S) = \emptyset.$
\end{proposition}

\begin{proof}
  Let $s\in\N,$ and an execution $w$.  We denote $T=geo(w_{\mid
    s}$. Consider a process $p\in\Pi$, for all process $q\neq p$ we
  have $p\causal\limits^{s}_w q$. Since $w$ is fair in $\Pi$, we can
  consider $r>s$ the time at which $p$ is influencing all $q$ from
  round $s$. At his step, for all $q\neq p$, the barycentric
  coordinate of the vertex of $geo(w_{\mid r}(S)$ of colour $q$
  relative to the vertex of $geo(w_{\mid r}(S)$ of colour $p$ is
  strictly positive. This means that $geo(w_{\mid r}(S)$ does not
  intersect $\delta(T,\Pi\setminus{p})$.

  Since $w$ is fair in $\Pi$, we can repeat this argument for any
  $p\in\Pi$. We denote the $r^*$ the maximal such $r$  and since $\bigcup_{p\in\Pi}\delta(T,\Pi\setminus{p}) =
  \delta(T)$, we have that $\delta(geo(w_{\mid s})(S)) \cap
  geo(w_{\mid r^*})(S) = \emptyset.$
\end{proof}

\subsection{A Characterization of Geo-Equivalence}

We start by simple, but useful,  sufficient conditions about the size of geo-classes.
\begin{proposition}[$Fair(\Pi)$ is separable]\label{prop:faircard}
  Let $w\in IIS_n$, denote $\Sigma$ the geo-class of $w$. 
	If $w$ is $Fair$ on $\Pi$, then $\#\Sigma = 1$.
\end{proposition}
\begin{proof}
  Let $w'\in\Sigma$. We will show that $w'$ shares all prefixes of
  $w$.  Let $r\in\N$. From Prop.~\ref{prop:geoFair} and
  Lemma~\ref{lemma:convexSimplex}, we get that $geo(w)$ does not
  belong to the boundary of $geo(w_{\mid r})(S)$, nor to the boundary of $geo(w'_{\mid r})(S)$.  Assume that
  $w$ and $w'$ have not the same prefix of size $r$, that is
  $geo(w_{\mid r})(S) \neq geo(w'_{\mid r})(S)$.  From
  Prop.~\ref{prop:geoEquivalenceIntersection} $geo(w_{\mid r})(S)$,
  $geo(w'_{\mid r})(S)$ have to intersect (as simplices), and since
  they are different, they can intersect only on their boundary.  This
  means that $geo(w)$ would belong to the boundary, a contradiction.

  So they have the same prefixes and $w'=w$.
\end{proof}

\begin{proposition}[Infinite Cardinal]\label{prop:CardInfi}
  Let $n\geq 2$. Let $w$, $w'$ two distinct executions such that
  $geo(w) = geo(w')$ and there exist $s\in\N$ such that $\forall r>s
  \exists T geo(w_{\mid r})(S)\ \cap\ geo(w'_{\mid r})(S) = T$ with
  $T$ a simplex of dimension $k \leq n-2$. Then, the geo-equivalence
  class of $w$ is of infinite size.
\end{proposition}

\begin{proof}%
Let $w$, $w'$ two executions with $geo(w) = geo(w')$ and $\forall r
>s,\ geo(w_{\mid r})(S)\ \cap\ geo(w'_{\mid r})(S) = T$ with $T$ od
dimension $k \leq n-2$. Denote $P$ the colors of $T$. Since $k \leq
n-2$, we have at least $p_1\neq p_2 \in \Pi \setminus P$.
The suffix at length $s$ of $w$ is in $Solo(P)$.

Hence, for the processes in $P$, when running in $w$ or $w'$, it is
not possible to distinguish these 3 cases about the induced subgraph
by $\{p_1,p_2\}$ in the instant graphs : $p_1 \leftarrow p_2$, $p_1
\leftrightarrow p_2$ and $p_1 \rightarrow p_2$.

So $\forall r>s$, we have 3 possible ways of completing what is
happening on the induced subgraph by processes in $P$ in $G\in
Solo(P)$.  So we have infinitely many different executions, the
cardinality of the geo-class of $w$ is infinite.
\end{proof}

Let's consider the remaining cases.
Let $w\in IIS_n$, denote $\Sigma$ the geo-class of $w$. 

\begin{proposition}[Boundaries of $S^n$ are separable]
  If $w$ is $Fair$ on $\Pi\setminus \{p\}$ for some $p\in\Pi$ then $\#\Sigma = 1$.
\end{proposition}
\begin{proof}
  We denote $Q=\Pi\setminus\{p\}$. We apply Prop.~\ref{prop:faircard}
  for $n-1$ to $w'$ the restriction of $w$ to the set of processes $Q$ (this is
  possible by definition of $Fair$: no process of $Q$ receives message
  from outside of $Q$). Since $w'$ satisfies the condition for Prop.~\ref{prop:faircard}
   (by definition of Fair), which means that the geo-class of $w'$ is of size 1.

  Since there is only one unique way of completing an execution
  restricted to $Q$ to one in $Solo(Q)$ (adding $(q,p), \forall q\in Q$), we get that there is
  only $w$ in the equivalence class.
\end{proof}

A suffix of a word $w$ is \emph{strict} if it is not equal to $w$. 
\begin{proposition}
  If $w$ has only a strict suffix that is $Fair$ on $\Pi\setminus
  \{p\}$ for some $p\in\Pi$ then $\#\Sigma = 2$;
\end{proposition}
\begin{proof}
  We denote $Q=\Pi\setminus\{p\}$.  We
  can write $w=uav$ where $u\in IS_n^*$, $a\in IS_n$ and $v$ is Fair on $Q$ but $av$ is
  not. We can choose $u$ such that $u$ has the shortest length.

  We consider $w'$ such that $geo(w')=geo(w)$. Let $r$ be the length
  of $ua$. We denote by $T$ the facet of $geo(ua)(S^n)$ with colors
  $Q$. Since $v$ is Fair for $Q$, we can apply to $v_{\mid Q}$ the
  restriction of $v$ to $Q$ Prop.~\ref{prop:geoFair}. So $geo(w)$ is
  not on the boundaries of $T$ which means, from
  Prop.~\ref{prop:geoEquivalenceIntersection}, that either
  $geo(w'_{\mid r})(S^n)=geo(w_{\mid r})(S^n)$ either $geo(w'_{\mid
    r})(S^n)\cap geo(w_{\mid r})(S^n) = T$.

  In both cases, we can apply Prop.~\ref{prop:faircard} to $v'$ the
  restriction of $w$ to $Q$. Which means that there is only one
  restricted execution in $Q$. Since there is only one way to complete
  to $p$, there are as many elements in the class that simplices at
  round $r$ that include $T$.  Since we have a subdivision, we have
  exactly two simplices sharing the facet $T$.
  
  In the first case, this means that $w'_{\mid r}=ua$ and $w=w'$.

  In the second case, we have that $w'_{\mid r}=ub$ for some $b\neq
  a$. We remark that if $w'_{\mid r-1}\neq u$ this would
  contradict the minimality of $u$. Indeed, the prefixes of length
  $r-1$ are different, this means that $av$ is Fair for $Q$.
\end{proof}

Using these previous propositions, and remarking that for any $w$,
there exists $P$ such that $w$ has a suffix in $Fair(P)$, we can now
present our main result regarding the complete classification of
geo-equivalence classes.  Let $n\in\N$ and $\Sigma$ a geo-equivalence
class on $S^{n}$.  Then there are exactly 3 cardinals that are
possible for $\Sigma$ (only 2 when $n=1$, the case of
\cite{2generals-journal}):

  \begin{theorem}\label{thm:geoEquiv}
    Let $w\in IIS_n$, denote $\Sigma$ the geo-class of $w$. 
	\begin{itemize}
        \item[$\mathcal C_1$ :] If $w$ is $Fair$ on $\Pi$ or on
          $\Pi\setminus \{p\}$ for some $p\in\Pi$, then $\#\Sigma =
          1$;
	\item[$\mathcal C_2$ :] $w$ has only a strict suffix that is
          $Fair$ on $\Pi\setminus \{p\}$ for some $p\in\Pi$ then
          $\#\Sigma = 2$;
	\item[$\mathcal C_\infty$ :] otherwise $\Sigma$ is infinite.
	\end{itemize}
\end{theorem}

\section{The Set-Agreement Problem}
\label{sec:setag}
For all $n$, the \emph{set-agreement problem} is defined by the
following properties \cite{Lynch}. Given initial $init$ values in $[0,n]$, each process
outputs a value such that

\begin{description}
\item[Agreement] the size of the set of output values is at most $n$,
\item[Validity] the output values are initial values of some processes,
\item[Termination] All processes terminates.
\end{description}

We will consider in this part sub-IIS message adversaries $\ma$, that is
$\ma\subseteq IIS_n$.
It is well known that set-agreement is impossible to solve on $IIS_n$,
we prove the following characterization.
\begin{theorem}\label{th:carac}
  Let $\ma\subset IIS_n$. 
  It is  possible to solve Set-Agreement on \ma if and only if
  $\wr\ma\wr \neq |S^n|$.
\end{theorem}

\subsection{Impossibility Result}

On the impossibility side, we will prove a stronger version with
non-silent algorithms. An algorithm is said to be \emph{non-silent} if
it sends message forever. Here, this means that a process could have
decided a value while still participating in the algorithm.

\begin{theorem}\label{th:cn}
  Let $\ma\subset IIS_n$. If $\wr \ma\wr = \wr IIS_n\wr = |S^n|$ then
  it is not possible to solve Set-Agreement on \ma, even with a
  non-silent algorithm.
\end{theorem}

We will need the following definition from combinatorial topology.
\begin{definition}[Sperner Labelling]
  Consider a simplicial complex $C$ that is a
  subdivision of a chromatic simplex $(S,\chi)$.
  A labelling $\lambda: V(C)\longrightarrow\Pi$ is a \emph{Sperner
    labelling} if for all $x\in V(C)$, for all $\sigma\subset S$, we
  have that $x\in|\sigma|\Rightarrow \lambda(x)\in\chi(\sigma)$.
\end{definition}

\begin{lemma}[Sperner Lemma \cite{Sperner}] 
  Let a simplicial complex $C$ that is a subdivision of a
  chromatic simplex $(S,\chi)$ with Sperner labelling $\lambda$.
  Then there exists $\sigma\in C$, such that $\lambda(\sigma)=\Pi.$ 
\end{lemma}

  A simplex $\sigma$ with labelling using all $\Pi$ colors is called
  \emph{panchromatic}.

\begin{proof}[Proof of Theorem~\ref{th:cn}{}]
  By absurd, we assume there is a non-silent algorithm $\algo$ (in full
  information protocol form) solving
  set-agreement on $\ma$.  We run the algorithm on initial inputs
  $init(i)=i$. We translate the full
  information protocol to the chromatic average, non-silent form : the
  initial value of $i$ is $x_i^*$; when the
  decision value is given, we still compute and send the chromatic
  average forever.  We can also assume a ``normalized'' version of the
  algorithm : when a process receives a decision value from a
  neighbour, it will decide instantly on this value. Such a
  normalization does not impact the correctness of the algorithm since
  set-agreement is a colorless task.

  The proof will use the Sperner Lemma with labels obtained from the
  eventual decision value of the algorithm. However it is
  not possible to use directly the Sperner Lemma for the 
  ``full subdivision under \ma'' (which we won't define), since this subdivision could be infinite.
  The following proof will use König Lemma to get an equivalent statement.

  Given $t\geq0$, we consider $\chr^t(S^n)$ under our algorithm with
  initial values $init(i)=i$.  For any vertex, we define the following
  labelling $\lambda_t$: if the process $i$ has not terminated at time
  $t$ with state $x\in V(\chr^t(S^n))$, then the Sperner label
  $\lambda_t(x) = i$, otherwise it is the decided value.
  Since the decided value depends only on the local state, the label
  of a vertex at time $t$ is independent of the execution leading to
  it.
  The goal of the following is to show that there is an entire
  geo-equivalence class that does not belongs to $\ma$. 
\medskip  

  By Integrity property, we have that the value decided on a face of
  $S^n$ of processes $i_1,\dots,i_n$, ie for
  $Solo(i_1,\dots,i_n)^\omega$ are taken in $i_1,\dots,i_n$.
  From Prop.~\ref{prop:solomegaboundary}, at any $t$, this labelling defines therefore a Sperner labelling of a
  (chromatic) subdivision of $S$.

  We consider the set $\mathcal S$ of all simplices $S$ of dimension $n$
  of $\chr^t(S^n),$ for all $t$. For a given $t$, from Sperner Lemma, there is at
  least a simplex of $\chr^t(S^n),$ that is panchromatic. There is
  therefore an infinite number of simplices $S$ that are panchromatic
  in $\mathcal S$. We consider now $\mathcal T\subset\mathcal S$, the
  set of simplices $T\in\mathcal S$ such that there is an infinite
  number of panchromatic simplex $S$ such that $|S|\subset|T|$. Note
  that $T$ needs not be panchromatic. 
  Since the number of simplices of $\chr^t(S^n)$ is finite for a given
  $t$, there is at least one simplex of $\chr^t(S^n)$ that is in
  $\mathcal T$. Therefore the set $\mathcal T$ is infinite.
  
  We build a rooted-tree structure over $\mathcal T$ : the root is
  $S^n$ (indeed it is in $\mathcal T$), the parent-child relationship
  between $T$ and $T'$ is defined when $T'\in\chr(T)$. We have an
  infinite tree with finite branching. By König Lemma, we have an
  infinite simple path from the root. We denote $T_t$ the vertex at
  level $t$ of this path. We have $|T_{t+1}|\subset |T_t|$ and
  $(T_t)_{t\in\N}$ converges (same argument as the end of
  Section~\ref{sec:converge}) to some $y\in |S^n|$. The increasing
  prefixes corresponding to $T_t$ define an execution $w$ of $IIS_n$.

  We will now consider two diffrent cases, not on the fact whether or
  not, $w\in\ma$, but on the result of $\algo$ on execution $w$.
  
  For first case, assume that algorithm $\algo$ has eventually decided on all processes on run
  $w$ at some time $t_0$. Since it could be that $w\notin\ma$, we cannot conclude yet.
  But since all processes have decided, they do not change their label in
  subsequent steps. By definition, $T_{t_0}=geo(w_{|t_0}(S^n)$ contains an
  infinite number of panchromatic simplices, \ie at least one.
  So the simplex $geo(w_{|t_0}(S^n)$ is panchromatic. Hence any run with prefix
  $w_{|t_0}$ cannot be in \ma, since \algo solves set-agreement on \ma.
  Therefore $w'=w_{|t_0}\mathcal K_\Pi^\omega$ ( where $\mathcal K_\Pi$ is the
  complete graph), is a fair execution that does
  not belong to $\ma$. Its entire geo-equivalence class, which is a
  singleton, is not in $\ma$.

  The second case is when algorithm $\algo$ does not eventually decide on all processes on run
  $w$.  Therefore $w\notin\ma$.
Now we show that all elements $w'$ of the geo-class of $w$ are also not in \ma.
Assume otherwise, then \algo halts on $w'$. By Prop.~\ref{prop:geoEquivalenceIntersection}, at
any $t$, the simplex corresponding to $w'_{\mid t}$ intersects $T_t$
on a simplex of smaller dimension whose geometrization contains $y$.
Consider $t_0$ such that the execution has decided at this round for
$w'$. Consider now $T_{t_0+1}$, it intersects the decided simplex of $\chr^{t_0}(S^n)$ corresponding to
$w'$, which means that the processes corresponding to the intersection were solo in
$w'(t_0+1)$ and in $w(t_0+1)$.  When a process does not belong to a
set of solo processes of the round, it receives all their values.
So by normalization property of algorithm \algo,
this means that in $T_{t_0+1}$, all processe have decided. A
contradiction with the fact that $\algo$ does not decide on all processes on run
  $w$. 
\end{proof}

This impossibility result means that there are many strict subsets
\ma of $IIS_n$ where it is impossible to solve set-agreement,
including cases where $IIS_n\setminus\ma$ is of infinite size.

\subsection{Algorithms for Set-Agreement}
In this section, we consider message adversaries \ma that are of the
form $IIS_n\setminus geo^{-1}(y)$ for a given $y\in| S^n|$. We note
$w\in IIS_n$, such that $geo(w)=y$. We have $w\notin\ma$.  In other
words, $\ma = IIS_n\setminus \mathcal C$, where $\mathcal C =
geo^{-1}(geo(w))$ is the equivalence class of $w$.
We also denote $\sigma_y(r)$ the simplex $geo(w_{\mid r})(S^n)$.

\subsubsection{From Sperner Lemma to Set-Agreement Algorithm}
Remark that the protocol complex at time $r$ is exactly $\chr^r(S^n)$,
there is no hole ``appearing'' in finite time for such \ma.
From Sperner Lemma, any Sperner labelling of a subdivision of $S^n$ admits at
least one simplex that is panchromatic. %
In order to solve set-agreement, the idea of
Algorithm~\ref{algo:set} is to try to confine the panchromatic, 
problematic but unavoidable, simplex of $\chr^t(S^n)$ to
$\sigma_y(r)$.
Since the geo-class of $w$ is not
in $\ma$, any execution will eventually diverge from $\sigma_y(r)$ and end
in a non panchromatic simplex.
We now define a special case of Sperner labelling of the Standard Chromatic
Subdivision that admits exactly one given simplex that is panchromatic.

We consider the generic colored simplex $(S,\chi)$ where
$S=(x_0,\dots,x_n)$ and coloring function $\chi$, that could be different
from $\mathcal P$. We consider labellings of the complex $Chr(S)$.

\begin{definition}%
	Let $\tau \in Chr(S,\chi)$. $f:V(Chr(S,\chi)) \longrightarrow
        \Pi$ is a \emph{Sperner $\tau-$panlabelling} if :
        $f$ is a Sperner labelling of $(S,\chi)$;
        for all simplex $\sigma\in\chr(S)$, $f(\sigma) = \Pi$ if and
          only if $\sigma = \tau$.
\end{definition}

\begin{proposition}\label{prop:panlabel}
Let $\tau$ a maximal simplex of $Chr(S,\chi)$, there exists a
$\tau-$panlabelling $\lambda$ of $Chr(S,\chi)$.
\end{proposition}

This key proposition is proved in the next section. Denote $\lambda_\tau(S,\chi)$ such
a Sperner $\tau-$panlabelling of $\chr(S,\chi)$.

\medskip
Before stating the algorithm, we show how to construct 
a sequence of panlabellings for $\chr^r(S^n)$.  Let $r\in\N$, we denote
$\Psi_w(r)$ the following labelling defined by recurrence.

Intuitively, it is the following labelling. In $\chr^{r}(S^n)$, we have
$\sigma_y(r)$ that is panchromatic, all other simplices using at most
$n$ colors. In $\chr^{r+1}(S^n)$, we label vertices that do not belongs
to the subdivision of $\sigma_y(r)$ by the labels used at step $r$. In
vertices from $\chr \sigma_y(r)$, we use $\lambda_{\theta(w(r+1))}$ the Sperner
$\tau-$panlabelling associated with $\theta(w(r+1))$ to complete the labelling that
uses at most $n$ colors on a given simplex, except at $\sigma_y(r)$.
In order to simplify notation, we also note $\lambda_G$ the labelling $\lambda_{\theta(G)}$.
Of course, we apply $\lambda_{w(r+1)}$ using as input
(corner) colors, the colors from $\Psi_w(r)$. This way, on the
neighbours of $\sigma_y(r)$ the labelling is compatible.

We denote $\gamma_r(x)$ the
precursor of level $r$ of $x\in V(\chr^{r+1}(S^n))$, that is the vertex
of $V(\chr^{r}(S^n))$ from which $x$ is originating.
\begin{definition}
We set $\Psi_w(1)(x) = \lambda_{w(1)}(S^n,\mathcal P)(x)$ for all
$x\in V(\chr^r(S^n))$, and for $r\in\N^*$
\begin{eqnarray*}
  \Psi_w(r+1)(p)&=&\Psi_w(r)(\gamma_r(x)) \mbox{ if } x\notin | geo(w_{\mid r})(S^n)|\\
  && \lambda_{w(r+1)}(\Psi_w(r)(\sigma_y(r))(x) \mbox{ if } x\in | geo(w_{\mid r})(S^n)|\\
\end{eqnarray*}
\end{definition}

\begin{proposition}
For all $r$, $\Psi_w(r)$ is a Sperner $\tau-$panlabelling of $\chr^r(S^n)$ for $\sigma_y(r)$.
\end{proposition}
\begin{proof}
  The proof is done by recurrence. The case $r=1$ is Prop~\ref{prop:panlabel}.
  Assume that $\Psi_w(r)$ is a Sperner $\sigma_y(r)-$panlabelling of $\chr^r(S^n)$.

  Consider now $\Psi_w(r+1)$ for $\chr^{r+1}(S^n)$. By construction
  and recurrence assumption, panchromatic simplices can only lay in
  $|\sigma_y(r)|$. Since $\lambda_{w(r+1)}$ is a Sperner
  panlabelling and that the corner colors for $\sigma_y$
  are taken from $\Psi_w(r)$, we have that $\sigma_y(r)$ is
  the only panchromatic simplex of $\chr^{r+1}(S^n)$.
\end{proof}

\begin{algorithm}[t]
  $x \leftarrow x_i^*$;
  $r\leftarrow 0$\;
  \Loop{while $x\in V(geo(w_{|r})(S^n)$\label{loop}}{
    $r\leftarrow r+1$\;
    $Send((i,x))$\;
    $V\leftarrow$\texttt{Receive()} // set of all received messages\; %
    $d\leftarrow sizeof(V) - 1$ // $i$ receives $d+1$ messages \; 
    $x = \frac{1-\frac{d}{2d+1}}{d+1}x + \sum_{(j,x_j)\in V, j\neq i}\frac{1+\frac{1}{2d+1}}{d+1}x_j$\;
  }
  Output : $\Psi_w(r)(x)$\;
  \caption{Algorithm $\algo_w$ for process $i$\label{algo:set}.}
\end{algorithm}

We now prove the correctness of $\algo_w$ presented in Algorithm~\ref{algo:set}.
Consider an execution $v\in\ma$.
For Termination: since elements of the geo-class of $w$ are not in
$\ma$, there exists a round $r$ at which
$v_{\mid r}\neq w'_{\mid r}$ for all $w'\in geo^{-1}(geo(w)$, \ie the
conditional at line~\ref{loop} is false
for all processes and the algorithm is terminating.
For Agreement: when terminating at round $r$, $i$ is not in $\sigma_y(r)$, by loop conditional, so since $\Psi_w(r)$ is only panchromatic on $\sigma_y(r)$, the number of decided values is less than $n$.
Integrity comes from the fact that $\Psi_w(r)$ is a Sperner labelling.

\subsubsection{Sperner Panlabellings of the Standard Chromatic Subdivision}
In this section, $n$ is fixed. We show how to construct a Sperner
panlabelling of the standard chromatic subdivision.
We consider the generic colored simplex $(S,\chi)$ where
$S=(x_0,\dots,x_n)$ and coloring function $\chi$, that could be different
from $\mathcal P$. We consider labellings of the colored complex $Chr(S,\chi)$.

We show the following combinatorial result about Sperner labellings.
\begin{theorem}
Let $\tau$ be a maximal simplex of $Chr(S,\chi)$, then there exists a
$\tau-$panlabelling $\lambda$ of $Chr(S,\chi)$.
\end{theorem}

We start by some definitions related to proving the above theorem.
It is possible to associate to any simplex $\sigma$ of $Chr(S)$ a
pre-order $\succ$ on $\Pi$ that corresponds to the associate graph
$\Theta(\sigma)$ :
$i\succ j$ when $(i,j)\in A(\Theta(\sigma))$.  We call equivalence classes for
$\Theta(\sigma)$, the classes of the equivalence relation defined by
$i\succ j \wedge j\succ i$. It corresponds actually to the strongly connected
components of the directed graph $\Theta(\sigma)$.

We define the \emph{process view} of a point.
This is the color of points in the view $V$ of vertex $(i,V)$ of the standard chromatic subdivision.
\begin{definition}[Process View]
The \emph{process view} of point $x = (\chi(x),V) \in V(\chr(S,\chi))$ is defined by~:
$V_x = \{\chi(y) | y\in V\}.$%
\end{definition}

For $\tau\in\chr(S,\chi)$, we also define the \emph{process view
relative to $\tau$} of a process $p$, denoted $V_p^\tau$. It is the
process view of the point of $\tau$ whose color is $p$.  It is linked
to pre-order $\succ$ : we have $V_p^\tau = \{q \mid q \succ p\}$.

Let $\tau$ be a fixed maximal simplex of $Chr(S)$.
We show how to construct a $\tau-$panlabelling. We choose a
permutation $\varphi$ on $\Pi$ such that it defines circular
permutations on the equivalent classes of $\Theta(\tau)$. Let $p\in\Pi$,
given $W\subset V_p^\tau$ such that $p\in W$, we denote by
$min^*(p,W) = \min\{i\in\N^*\mid \varphi^{i}(p)\in W\}$. Note that
since $\varphi$ is a permutation, there exists $j>0$ such that
$\varphi^j(p) = p$, and since $p\in W$, the minimum is taken over a
non-empty set. Finally we set $\varphi^*(p,W) = \varphi^{min^*(p,W)}(p).$
This is the first point of $W$ that is in the orbit of $p$ in $\varphi$.

\begin{definition}
We define  $\lambda_\tau : V(Chr(S)) \rightarrow V(S)$, for $x\in
V(Chr(S))$, we set

$\lambda_\tau(x) = 
\begin{cases} 
q &\text{ if } \exists q \in V_x \text{ and } q \notin V_{\chi(x)}^{\tau}\\
\varphi^*(\chi(x),V_x\cap V_{\chi(x)}^\tau) & \text{otherwise.}
\end{cases}$

\end{definition}

Intuitively, for a given vertex of $Chr(S)$ with view $V$, if the
process sees an other process $q$ than in $\tau$, then it is labelled
by this $q$, otherwise it will choose the first process in the
circular orbit of $\varphi$ that is in its view.

\begin{proposition}
The labelling $\lambda_\tau$  is a $\tau-$panlabelling.
\end{proposition}

\begin{proof}
First we show that it is indeed a Sperner labelling. In both cases of
the definition, $\lambda_\tau(x)$ belongs to $V_x$. For $x\in
V(Chr(S))$, for $\sigma\subset S$, $x\in |\sigma|$, with $\sigma$
minimum for this property, means that the presentation of $x$ is
$\Phi(x) = (i,\sigma)$ for some $x_i$ such that $x_i\in V(\sigma)$ and
$\chi(x_i)=i$.

Now we show that the only panchromatic simplex is $\tau$.  By
construction, with $x\in V(\tau)$, $\varphi^*(\chi(x),V_x) =
\varphi(\chi(x))$ since in this case $V_x = V_{\chi(x)}^\tau$. So $\tau$ is panchromatic through $\lambda_\tau$.

\medskip
Now we consider $\sigma \neq \tau$. We have two possible 
cases :

\begin{enumerate}
\item\label{dessous} $\exists x\in V(\sigma),q \in V_x, q \notin V_{\chi(x)}^{\tau}$,
\item\label{riendessous} $\forall x\in V(\sigma), V_x \subseteq
  V_{\chi(x)}^{\tau}$.%
\end{enumerate}

We start with the first case, we denote by $C$ the highest, for $\succ$ in $\sigma$, class
such there is $x$ in $C$ satisfying the clause (\ref{dessous}).
We show that $\# \lambda_\tau(C)\cap C < \#
C$, where $\#$ denotes the cardinal of a set.
By definition of $C$, $\lambda^{-1}_\tau(C)\subseteq C$. Since
$\lambda_\tau(x)\notin C$, this means that $\# \lambda_\tau(C)\cap C
\neq \# C$. By assumption all classes $C'$ that are higher than $C$ choose colors in $C'$,
so $\sigma$ is not panchromatic under $\lambda_\tau$.

Now, we assume we do not have case~(\ref{dessous}), this means that $\forall x\in V(\sigma),
\lambda_\tau(x) = \varphi^*(\chi(x),V_x)$.  Since $\sigma \neq \tau$,
there exists $x\in V(\sigma), V_x \varsubsetneq V_{\chi(x)}^{\tau}$.
We choose the lowest such $x$ for $\succ$ in $\tau$. %
We consider $C_x$ the class of $x$ in $\sigma$. We show that $\# \lambda_\tau(C_x) < \# C_x$.

We denote $C_x^\tau$ the class of color $\chi(x)$ in $\tau$. First we show that
$C_x\subseteq C_x^\tau$.  Indeed, assume there is $y\in C_x$ such that
$y\notin C_x^\tau$. Since the view of elements of the same class are
the same, this means that $\chi(x)\in V_y $ and $y$ would satisfy
property \ref{dessous}. A contradiction to the case we are considering.
And this is true for all $y\in C_x$.

Now we show $C_x\subsetneq C_x^\tau$. We have $V_x \subsetneq
V_{\chi(x)}^{\tau}$. Let $y\in V_{\chi(x)}^\tau\setminus V_x$. If $y\notin C_x^\tau$,
by the same previous argument, we get a contradiction. Hence $y\in
C_x^\tau$ and therefore $C_x\varsubsetneq C_x^\tau$.

We denote $p=\varphi^*(\chi(x),C_x^\tau\setminus
C_x)$.  We note $p' = \varphi^{-1}(p)$. We have by definition of
$\varphi^*(.,C_x^\tau\setminus
C_x)$, that $p'\in C_x$, therefore $C_{\chi^{-1}_{\mid
    \sigma}(p')} = C_{x}$. Now we set $p'' = \varphi^*(p',C_x)$. The
color $p''$ has at least two predecessors in the labelling : $p'$ by
construction (since $x$ was choosen the lowest for $\succ$ then
$V_{\chi^{-1}_{\mid \sigma}(p')}=V^\tau_{p'}$) and $p''' =
\varphi^{-1}(p'')$ which is not $p'$ since $\varphi(p')=p\notin C_x$.

So $\# \lambda_\tau(\chi^{-1}_{\mid \sigma}(V_x^\tau)) < \# V_x^\tau$, and $\lambda_\tau(\sigma)\neq \Pi$.
\end{proof}

\subsubsection{Lower Bounds}
It is possible to use the
impossibility result to prove the following lower bound.
Algorithm~\ref{algo:set} is therefore optimal for fair $w$.%
\begin{theorem}
  Let $\algo$ be an algorithm that solves set-agreement on $\ma=
  IIS_n\setminus geo^{-1}(geo(w))$ with $w\in IIS_n$. Then, for any execution $v$, $t\in\N$, such
  that $v_{\mid t} = w'_{\mid t}$ for some $w'\in geo^{-1}(geo(w))$, \algo has not terminated at $t$.
\end{theorem}

\begin{proof}
  Suppose \algo has decided on all process at $t$, with $v_{\mid t} = w'_{\mid t}$ for some $w'\in geo^{-1}(geo(w))$.
  So $\algo$ solve set-agreement on
  $w'$. A contradiction with Th.~\ref{th:cn} since $\wr \ma \cup\{w'\}\wr = |S^n|$.
\end{proof}

\section{Conclusion and Open Questions}
\label{sec:ccl}
In this note, we have presented how to construct a topology directly
on the set of executions of $IIS_n$ the Iterated Immediate Snapshot
message adversary. Though this is not a simple
textbook topology as usual, since there are non-separable points,
the characterization we presented enables to fully understand it. As a
first application, we were able to characterize for the first time general subsets
of $IIS_n$ where set-agreement is solvable.  We also believe
this new approach could be successfully applied to other
distributed tasks and distributed models.

Another interesting open question would be to compare the geometrization topology 
to the knowledge-based ones defined in
\cite{consensus-epistemo}.

\end{document}